\documentclass[11pt,a4paper]{article}

\usepackage[latin1]{inputenc}
\usepackage{graphicx}

\usepackage[top=3cm, bottom=3cm, left=2cm, right=2cm]{geometry} 

\begin{document}

\title{Long-time stable HTSC DC-SQUID gradiometers with silicon dioxide passivation for measurements with superconducting flux transformers}

\noindent\bfseries \LARGE Long-time stable HTSC DC-SQUID gradiometers with silicon dioxide passivation for measurements with superconducting flux transformers

\vspace{1cm}
\mdseries
\small\noindent
P Seidel$^1$, C Becker$^1$, A Steppke$^1$, M Buettner$^1$, H Schneidewind$^2$, V Grosse$^1$, G Zieger$^1$ and F Schmidl$^1$

\vspace{1cm}
\noindent
$^1$ Institut für Festkörperphysik, Friedrich-Schiller-Universität Jena, Helmholtzweg 5,
D-07743 Jena, Germany

\noindent $^2$ Institut für Photonische Technologien (IPHT) e.V., Albert-Einstein-Straße 9, D-07745 Jena,
Germany

\noindent paul.seidel@uni-jena.de


        \begin{abstract}
        In applications for high-T$_c$ superconducting DC-SQUIDs such as
        biomagnetism, nondestructive evaluation and the relaxation of
        magnetic nanoparticles, it is important to maintain reliable sensor
        performance over an extended time period. We designed and produced
        DC-SQUID gradiometers based on YBa$_2$Cu$_3$O$_{7-x}$ (YBCO) thin films
        which are inductively coupled to a flux transformer to achieve a
        higher sensitivity. The gradiometers are protected against ambient
        atmosphere and humidity by SiO$_2$ and amorphous YBCO layers.

        The noise properties of the sensor in flip-chip configuration,
        especially in unshielded environments, are shown. We present a
        comparison of Tl$_2$Ba$_2$CaCu$_2$O$_{8+x}$ (TBCCO) thin films on buffered sapphire
        or LaAlO$_3$ substrates for the flux transformer in shielded and
        unshielded environments. We reach a low white field gradient noise of
        $72~\textrm{fT/(cm}\cdot \textrm{Hz})$ with the TBCCO on LaAlO$_3$ flux transformer. 
        The electric properties of the gradiometers (critical current $I_C$, normal state resistance $R_N$ and the
        transfer function $V_\Phi$) were measured over a period of one year
        and do not show significant signs of degradation.
        \end{abstract}


\normalsize

\section{Introduction}
Long-term sensor stability is a requirement for the application of
high temperature superconducting DC-SQUID gradiometers. Changes in
the electrical characteristics of sensors lead to decreased
performance and a higher cost of the final system. The most
sensitive areas of the gradiometer are the Josephson junctions,
which we create using a single layer bicrystal technology. An
overview of the properties of grain boundary Josephson junctions is
given in \cite{Gross2005}. A possible cause for degradation of
superconducting thin films is the diffusion of oxygen in the YBCO
layer. In monocrystalline films the oxygen transport occurs along
the a-b plane of the crystals \cite{Aarnink1992}. At growth defects
or grain boundaries a significant oxygen diffusion appears along the
c-axis of the epitaxially grown film \cite{kittelberger1998, Tian1999}.
Therefore a passivation at the sides and on top of the
superconducting thin film is necessary to reduce diffusion. We
developed a process to compensate for height differences during
fabrication and protect our sensors using different materials to
achieve stability over a period of more than one year.

\section{Fabrication}
\begin{figure}
\begin{center}
  \includegraphics[width=7.5 cm]{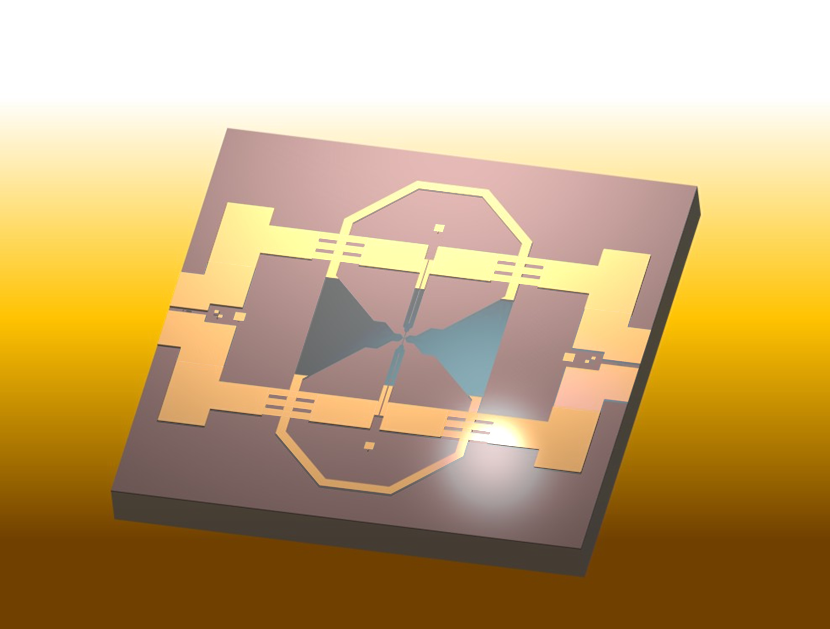}
  \caption{Schematic of the planar DC-SQUID gradiometer}
  \label{layout}
\end{center}
\end{figure}

The sensors described here are galvanically coupled DC-SQUID
gradiometers (for layout see figure \ref{layout}). The YBCO thin
films having a thickness of $150~\textrm{nm}$ are grown along the
c-axis on SrTiO$_3$ bicrystal substrates with pulsed laser
deposition \cite{Seidel2005}. The thin films are patterned with
Ar-ion beam etching and planarized {\it in situ} via the deposition
of $150~\textrm{nm}$ amorphous YBCO with RF-sputtering. For repeated
measurements with different combinations of readout gradiometers
and flux transformers the sensors need to be protected against
mechanical damage and require high-quality bond contacts. The direct
contact between readout gradiometer and flux transformer in the
flip-chip package can lead to damages in the thin films which
influence the electrical properties of the system. A protective
layer is also necessary to reduce diffusion of oxygen and to avoid
contact between the superconducting thin film and environmental
humidity. In our current setup we use SiO$_2$ for this layer on the
top of the YBCO film. Further details regarding the thin film system
is given in \cite{Seidel2007}.

\begin{figure}
\begin{center}
  \includegraphics[width=7.5 cm]{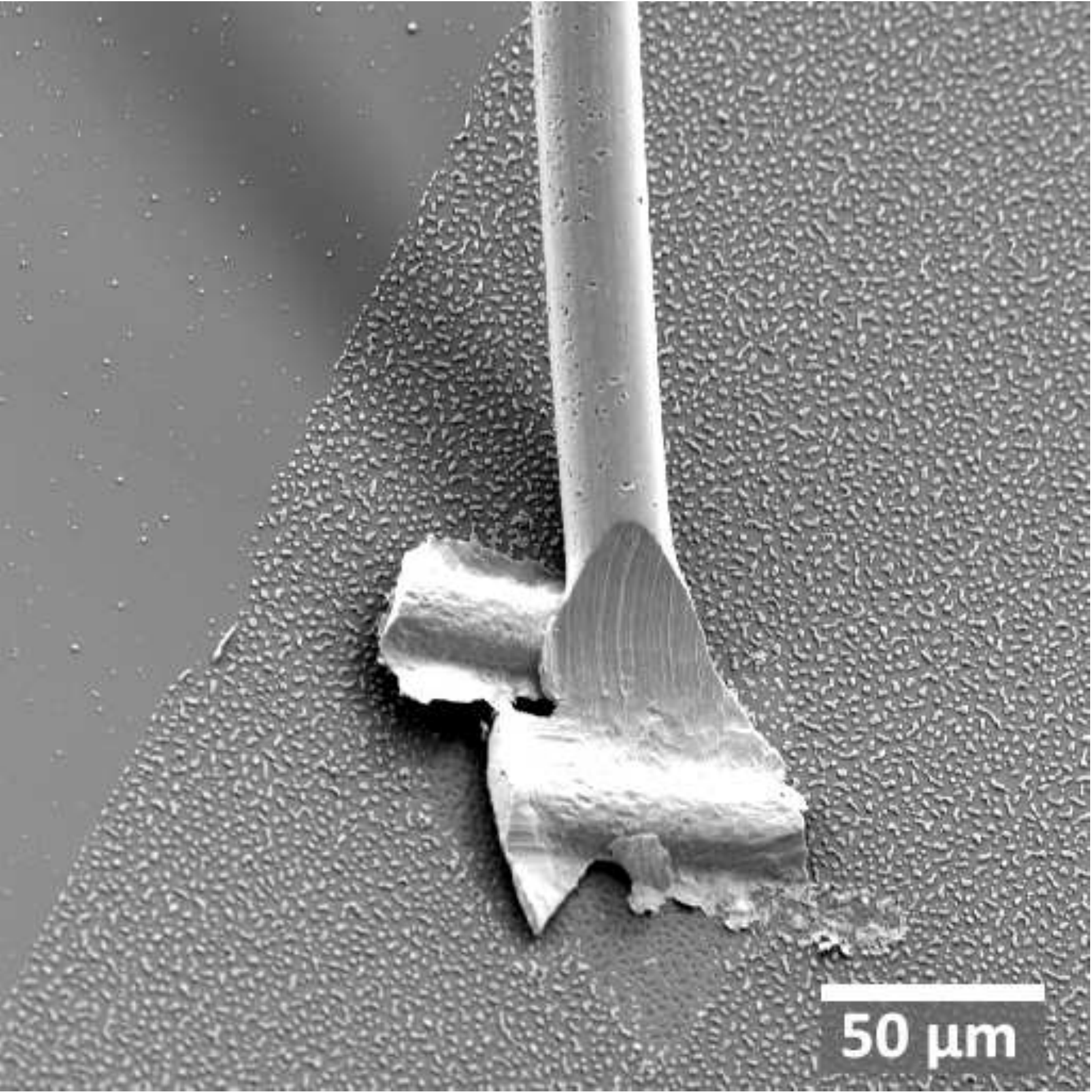}\\
  \caption{SEM image of a bonding pad with attached bond wire. Gold clusters provide good adhesion with low contact resistance to the YBCO film.}\label{gold}
\end{center}
\end{figure}

We developed a new technique to prepare the the bonding pads of the
sample. The aim was to improve the adhesion of the bond wire and to
lower the contact resistance. Therefore, prior the YBCO deposition,
we grow a $50~\textrm{nm}$ thin gold layer on the substrate using a
lift-off technique. During heating of the substrate to the
deposition temperature of 750°C the gold layer clusters to small
islands with a diameter of $1-2~\mu \textrm{m}$ (see figure
\ref{gold}). Between these islands the YBCO growths well c-oriented
with a rocking-curve width of less than 0.3° and a critical
temperature higher than $86~\textrm{K}$. The contact resistance
of an average bonding pad was lowered from $160~\textrm{m}\Omega$
(without gold clusters) to $<4~\textrm{m}\Omega$ (with gold clusters).

\begin{figure}
\centering
  \includegraphics[width=7.5 cm]{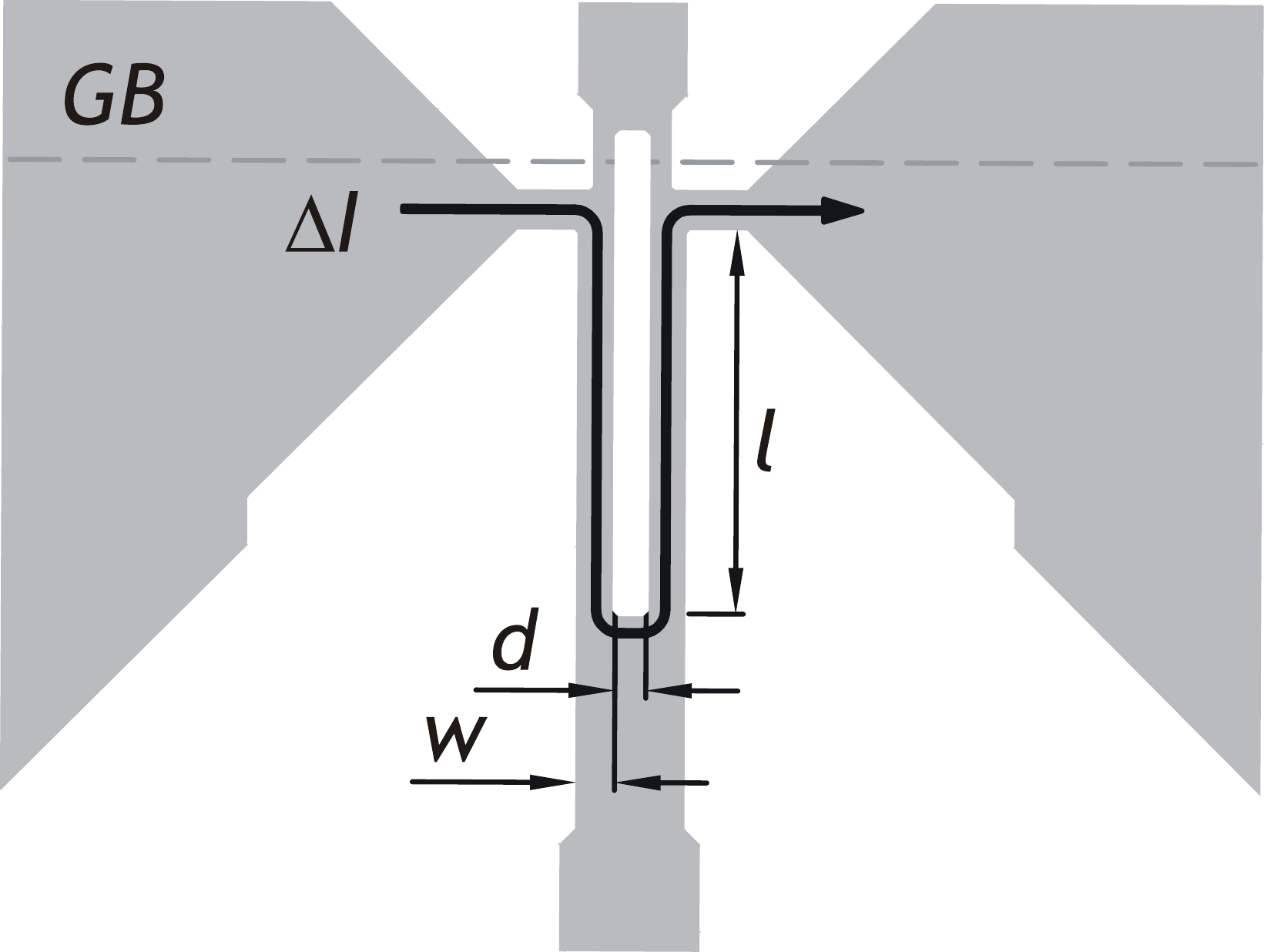}
  \caption{Inner part of the DC-SQUID gradiometer with loop length $l$, width of the superconducting lines $w$, inner loop width $d$ and grain boundary (GB). The difference in magnetic flux is galvanically coupled into the SQUID loop with the current $\Delta I = \Delta I_F \cdot K$, with the difference current in the flux transformer $\Delta I_F$ and the coupling constant $K$.}
  \label{layoutInnerPart}
\end{figure}

The SQUID consists of two Josephson junctions with a width of $3~\mu \textrm{m}$ and a galvanic coupling to the antenna structures. The pickup loops of the antenna structure are used in a gradiometric
configuration. An external magnetic field gradient leads to a difference in magnetic flux in the antenna structures. This difference is galvanically coupled into the
SQUID with the current $\Delta I$ (see figure \ref{layoutInnerPart}) and an inductance of the incoupling line of the SQUID loop $L_M$. The length of the SQUID loop is either $l = 50~\mu \textrm{m}$ or
in a second layout $l = 70~\mu \textrm{m}$ (see figure \ref{layoutInnerPart}), the width of the incoupling lines is $5~\mu
\textrm{m}$ and the inner hole area is $4~\mu \textrm{m}$ wide. 
In former investigations we have shown that
for a film thickness of about $150~\textrm{nm}$ the inductance $L_M$ of a DC-SQUID with a width of the
incoupling lines of $5~\mu \textrm{m}$ and an inner area of $3~\mu
\textrm{m}$ can be estimated by a factor of $1.05~\textrm{pH/}\mu \textrm{m}$ times
the loop length. The decrease of the width of incoupling lines to
$4~\mu \textrm{m}$ and the increase of the width of the inner area
to $5~\mu \textrm{m}$ increase this factor to a value of about
$1.8~\textrm{pH/}\mu \textrm{m}$ \cite{Schmidl1999}. 
With these specifications we estimate the coupling inductance of the DC-SQUIDs in our
gradiometers to be in the range of $50~\textrm{pH}$ in the ideal case, up to $90~\textrm{pH}$ for a loop length of $50~\mu \textrm{m}$. The exact value depends on the accuracy of the photolithography and patterning
process. To determine the coupling inductance we usually measure the
flux modulation of the DC-SQUID with a current directly coupled into
the SQUID loop \cite{Wunderlich1999}. This precise method is
only possible after disconnecting the antenna structures so that these
measurements can only be done after the long term experiments.

To optimize the coupling between the antenna of the readout
gradiometer and the antenna of the flux transformer we used
different widths $W_{ant}$ for the antenna of the readout gradiometer
($200~\mu \textrm{m}$, $400~\mu \textrm{m}$ and $600~\mu
\textrm{m}$, see figure \ref{layout})\cite{Peiselt2003}.

\begin{figure}
\centering
  \includegraphics[height=5cm]{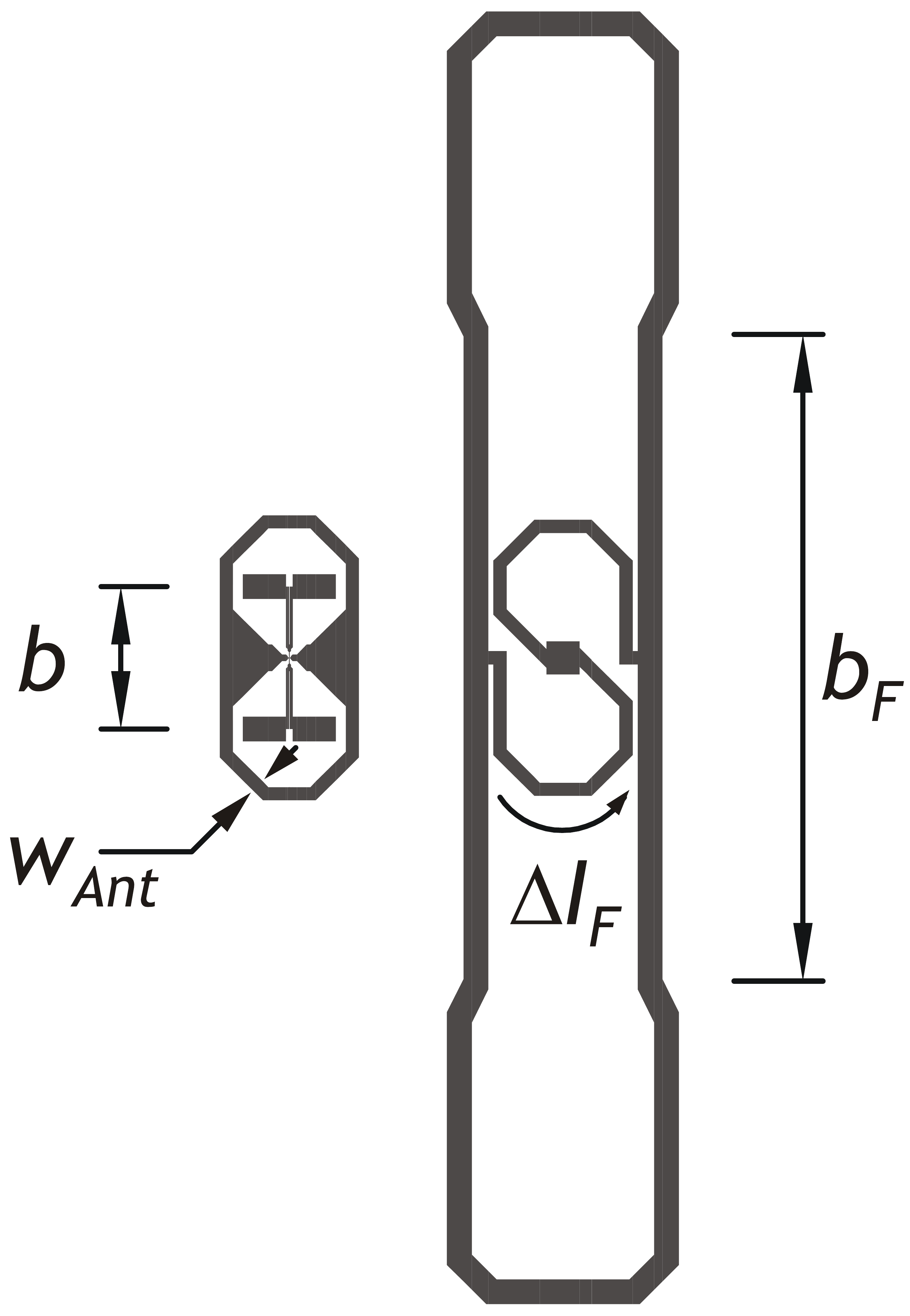}
  \caption{Flip chip configuration with readout gradiometer on the left side and flux transformer on the right side. The difference in screening currents $\Delta I_F$ is magnetically coupled into the readout gradiometer.}
  \label{layoutFlipChip}
\end{figure}

The flux transformer is prepared on a separate substrate (see figure \ref{layoutFlipChip}). We used
two different substrate materials and TBCCO as the superconducting
thin film material. The comparison between YBCO and TBCCO on different subtrate materials for the flux transformer has shown that TBCCO leads to the best sensor performance, i.e. the lowest white noise level, due to a better flux pinning in unshielded environment \cite{Seidel2007a}.
The size of the complete flux transformer is $8\times 40~\textrm{mm}^2$ and the baselength $b_f$ is
$19~\textrm{mm}$. After patterning by wet
etching $300~\textrm{nm}$ of SiO$_2$ are deposited on the flux
transformer as protection against chemical and mechanical influences.
The two pickup loops of the flux transformer are connected in
parallel. The current caused by a difference in magnetic flux is
inductively coupled to the readout gradiometer with an S-shaped
superconducting line, which is optimized to the dimensions of the
readout gradiometer.

\section{Electrical Characteristics}
\begin{figure}
\begin{center}
  \includegraphics[width=7.5 cm]{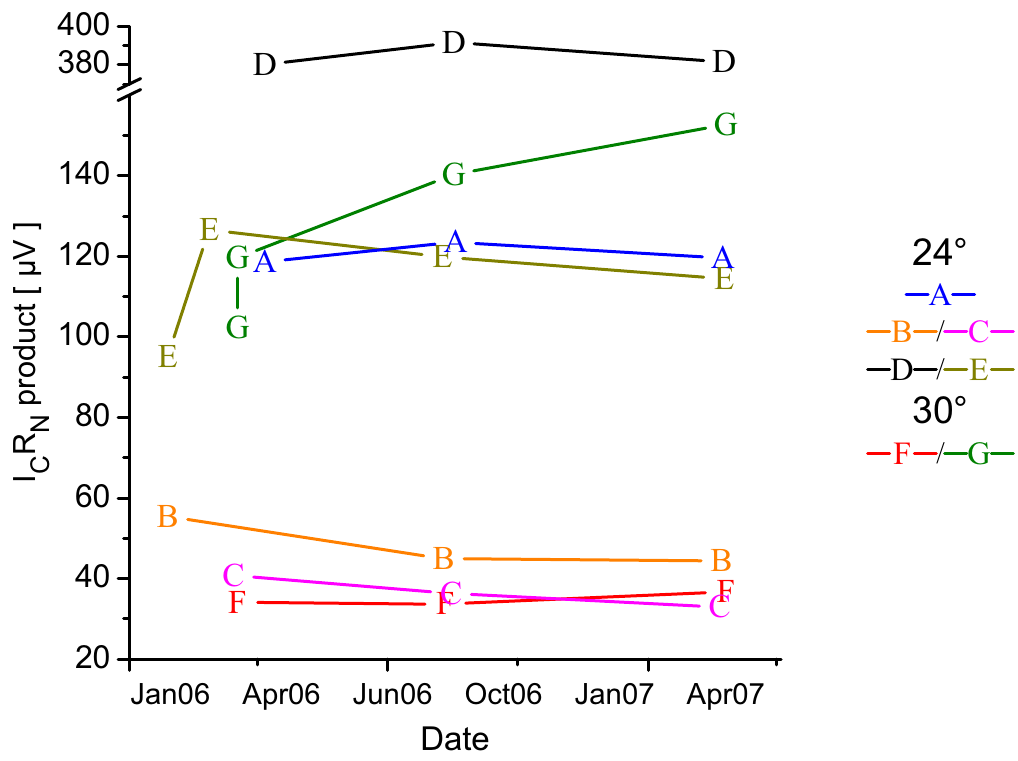}\\
  \caption{Time dependence of the critical current and the normal state resistance of seven readout gradiometers with grain boundary angles of 24° and 30°.}\label{icrn}
\end{center}
\end{figure}

\begin{figure}
\begin{center}
  \includegraphics[width=7.5 cm]{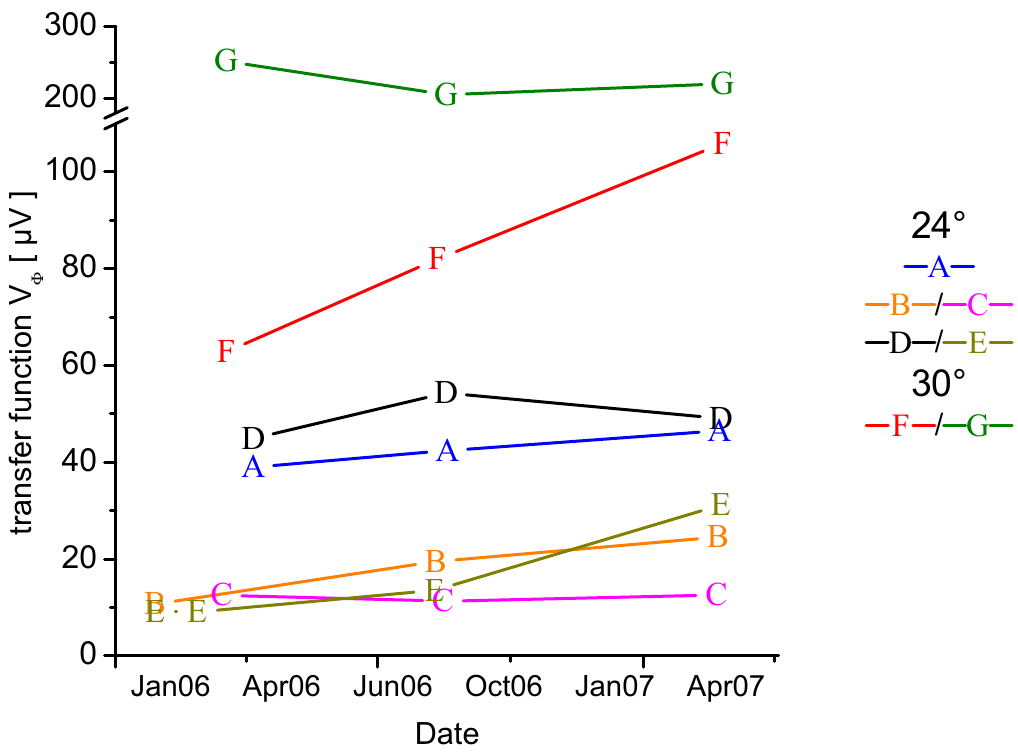}\\
  \caption{Time dependence of the transfer function of the readout gradiometers in figure \ref{icrn}.}\label{transf}
\end{center}
\end{figure}
We focused our measurements on the electrical characteristics at
liquid nitrogen temperatures. The sensors were evaluated over a
period of one year in a magnetically shielded environment (two
layers of $\mu$-metal) to reduce the influence of external
disturbances. During this period each sensor was cooled and then
heated to room temperature ten times to simulate environmental
stress during the normal sensor lifetime. Between the measurements
the gradiometers and flux transformers were stored at room
temperature with a humidity of 40--45\%.

To measure their I-V characteristics, the DC-SQUID gradiometers were
biased with different currents and the resulting voltage signal was
amplified and recorded. With this information we calculated the critical
current $I_C$, normal state resistance $R_N$ and the $I_CR_N$
product (see figure \ref{icrn}). The $I_CR_N$ product allows to
compare different types of Josephson junctions independent of the
junction geometry. This parameter should change significantly if
degradation between the measurements occurs. To evaluate the
response to external magnetic fields we calculated the transfer
function of the sensors over time (see figure \ref{transf}). At a
fixed bias current the maximum voltage modulation  $V_{pp}$
is measured as a function of the current in a small copper coil
$I_M$. Assuming a sinusodial $V(I_M)$ characteristic the transfer
function $V_\Phi$ can be calculated as $V_\Phi = \pi \cdot V_{pp}$.

\section{Experimental Results}
An important improvement to realize sensors with a long lifetime was the usage of SiO$_2$ passivation layers and amorphous YBCO films. As a first step we examined whether there is a negative influence of the passivation on the superconducting properties of the sensors. With all tested sensors the SiO$_2$ thin film did not change the electrical characteristics and the dielectric influence is negligible compared to the influence of the substrate material (SrTiO$_3$).

The planarisation of the structures with amorphous YBCO deposited by hollow cathode sputtering does not lead to a large change of the superconducting properties of bicrystal Josephson junctions. In some cases we observed a slight increase of the critical current, compared to bicrystal junctions without planarisation. This can be caused by oxygen diffusion between the planarisation layer consisting of YBCO$_{7-x}$ with $x \approx 0$ and the crystalline thin film with $0.1 > x > 0$. An increase of the critical current is primarily observed for bicrystal angles of 30° and above.

During the the processing of the gradiometer, especially during the dry etching and photolithography steps a temperature increase of the thin film up to 100°C accelerates the diffusion of oxygen into the area of the Josephson junctions instead of out of this area as observed without amorphous YBCO planarisation thin films. To lower the rate of diffusion out of the antenna areas the substrate is cooled by liquid nitrogen during the dry etching procedure \cite{Schneidewind1995,Alff1992}.

The comparison among gradiometers (see figure \ref{icrn}) shows a
large spread of the $I_CR_N$ product due to different current densities
of the grain boundary junctions used for these experiments. All of
these junctions show the typical
scaling behaviour for grain boundary junctions \cite{Seidel2007a}. Compared to previous
experiments by other groups only the $I_CR_N$ value of $30~\mu\textrm{V}$ for the 30° grain boundary junction (curve F)
is relatively low \cite{Minotani1997}. The experimental investigations
of the long term stability shows no relevant changes or signs of
degradation in most cases (see figure \ref{icrn}). Only small
changes of the $I_CR_N$ product were measured. In all of these cases
the reason for this behaviour was a decrease of the respective
critical current. A decreasing $I_C$ does not automatically lead to
a decrease in the transfer function, as aging processes can change
$R_N$ at the same time.

The 24° and 30° grain boundary junctions used in our experiments
exhibit relatively high critical current densities. In combination
with a width of the SQUID structures in the range of $3~\mu
\textrm{m}$ the critical currents can be as high as $100~\mu
\textrm{A}$. With high critical currents the inductance parameter
$\beta_L$ can exceed the optimum value ($\leq 1$), which leads to a
decreasing transfer function \cite{Wunderlich1999,Wunderlich2000}. Higher $\beta_L$ values reduce the voltage modulation due to a screening of
the flux which is coupled into the SQUID. Therefore a slight decrease of the critical current results in an improvement of the transfer function (sample B and E). Such a decrease can be
caused by diffusion during the processing of the superconducting
thin film or by slower diffusion during normal operation of the
sensor.

In the case of increasing critical current (sample G) we see a reduction of the transfer function $V_\Phi$ in figure \ref{transf}. In our experiments the transfer function increased significantly for only one sample with a 30° grain
boundary (sample F); the reason for this is topic of future investigations. %
%
%

The fabricated gradiometers display a large spread in the transfer function with a minimum of $10~\mu \textrm{V}$. However, all sensors were able to operate in unshielded environment
with commercially available SQUID electronics (Magnicon, Philips).
\normalfont

In the next step we have measured various combinations of readout gradiometers and
flux transformers in shielded as well as in unshielded environment
\cite{Seidel2007a}.
The field gradient resolution $\sqrt{S_G}$ is proportional to the flux noise $\sqrt{S_\Phi}$ and depends on the effective area of the configuration $A_{ef\!f}$ and the base length $b$:

\begin{equation}
    \sqrt{S_G} = \frac{\sqrt{S_\Phi}}{b\cdot A_{ef\!f} }
    \label{eq:FieldGradientResolution}
\end{equation}

\begin{figure}
\begin{center}
  \includegraphics[width=7.5 cm]{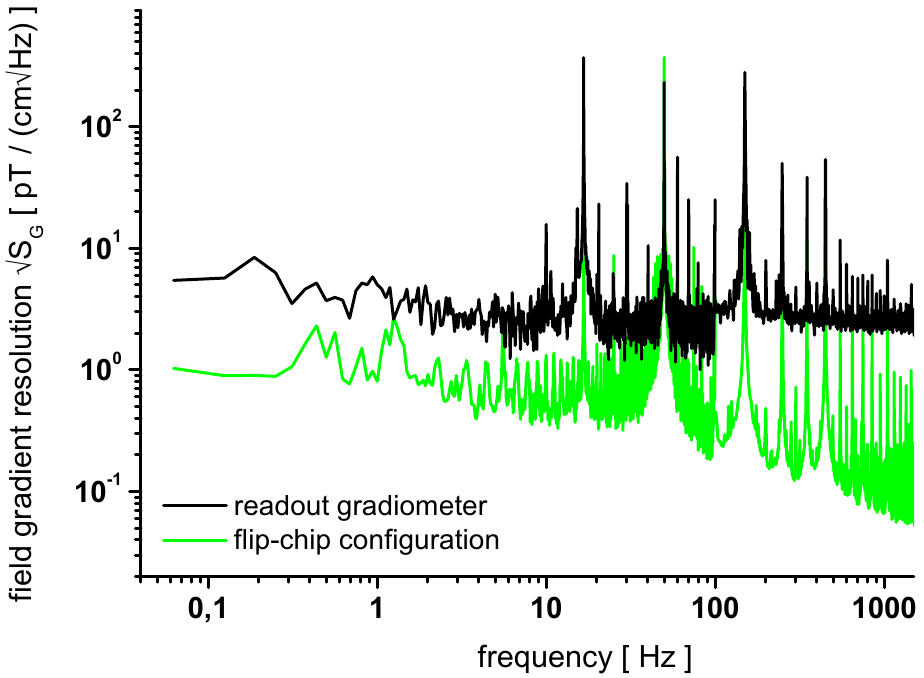}\\
  \caption{Field gradient resolution of the readout gradiometer without and with flux transformer.}\label{fieldresolution}
\end{center}
\end{figure}

The responsitivity of a gradiometer is defined as the product of effective area and base length.
Figure \ref{fieldresolution} shows as an example the increase of the field gradient resolution in unshielded environment
for the readout gradiometer (sample G) in comparison to the flip chip configuration. Due to the large flux transformer the flip chip configuration with a higher base length ($b_f \approx 1.9 \textrm{cm}$) and an increased effective area ($A_{ef\!f} \approx 0.22 \textrm{mm}^2$)

\begin{figure}
\begin{center}
  \includegraphics[width=12 cm]{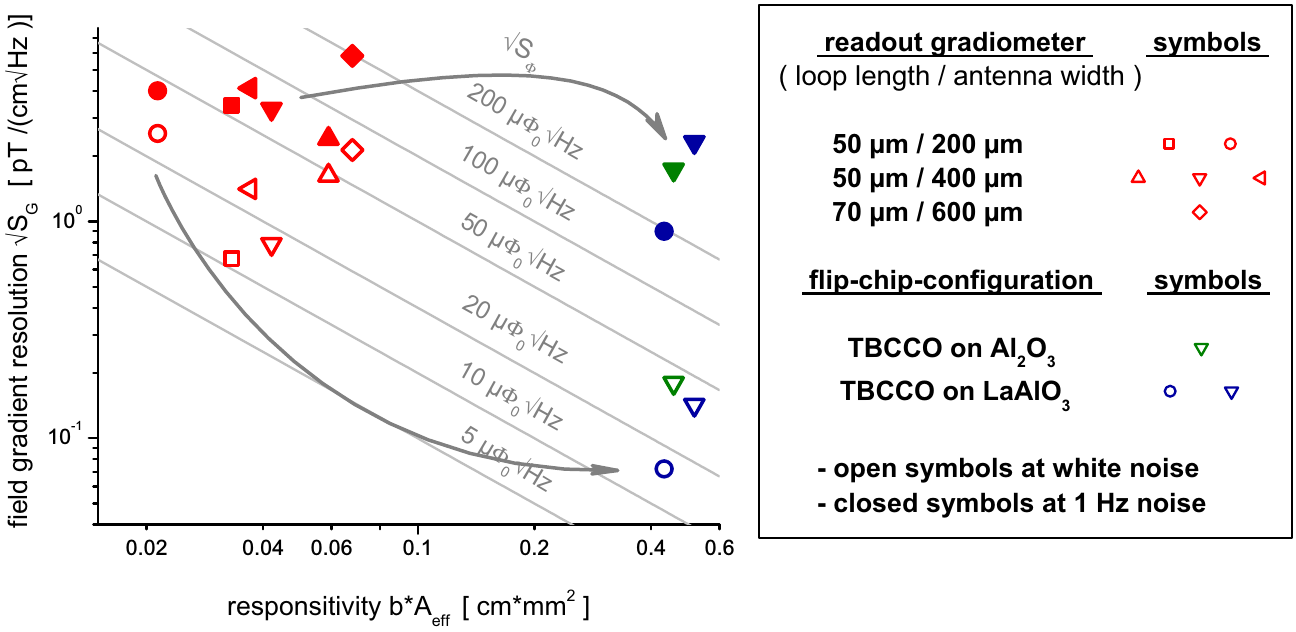}\\
  \caption{Field gradient resolution over responsitivity in magnetically unshielded environment with readout gradiometer on the left side and flip-chip configuration on the right side.}\label{noise}
\end{center}
\end{figure}

The field gradient resolution as a function of the responsitivity
for different read out gradiometers at 1 Hz and white noise region
is illustrated in figure \ref{noise}. By changing the width of
readout gradiometer antenna $w_{ant}$ the responsitivity can be
adjusted but the field gradient resolution is mostly unchanged. The
reason for this behaviour is the higher flux noise in unshielded
environment for larger antenna widths. Especially the largest
incoupling inductance $L_M$ (loop length $l = 70~\mu \textrm{m}$)
together with the largest antenna width results in the largest
responsitivity ($>0.06~\textrm{cm}\cdot \textrm{mm}^2$, $\diamond$),
but leads to a relatively low field gradient resolution in
unshielded environment, caused by the large flux noise of
$200~\mu\Phi / \sqrt{\textrm{Hz}}$ in comparison to the lower value
of about $50~\mu\Phi / \sqrt{\textrm{Hz}}$ for the other gradiometer
layouts.

This means that, for the development of sensors in flip chip configuration, 
it can be helpful to optimize the readout gradiometers in the
direction of low flux noise in unshielded environment, not in the
direction of higher responsitivity. As in the same figure shown we
realized the best value in the field gradient resolution with the
readout gradiometer with the lowest level in the flux noise ($ <
50~\mu\Phi / \sqrt{\textrm{Hz}}$, $l=50~\mu\textrm{m}$,
$200~\mu\textrm{m}$ line width, $\circ$) in unshielded environment
not with the sensor with the higher responsitivity
($0.04~\textrm{cm}\cdot \textrm{mm}^2$). If the readout gradiometer
alone has a large responsitivity ($A_{ef\!f}\cdot b$) the additional
flux transformer only improves the field gradient resolution by a
factor of two.

With a similar flux noise, it should be possible to reduce the field gradient resolution significantly due to the higher responsitivity of the combination of readout gradiometer and flux transformer. We are able to achieve a reduction by a factor of 5, corresponding to a field gradient resolution of $880~\textrm{fT/(cm} \sqrt{\textrm{Hz}})$ at frequencies around 1 Hz and
a reduction by a factor of 20 ($72~\textrm{fT/(cm} \sqrt{\textrm{Hz}})$) at higher frequencies, showing the effectiveness of this approach.

\section{Summary}
The developed and fabricated sensors were successfully protected by
a combination of an amorphous YBCO and SiO$_2$ layers against oxygen diffusion. After a
period of one year no significant signs of degradation were
detected.

With TBCCO flux transformers on LaAlO$_3$ substrates stacked with
readout gradiometers, we achieve a field gradient noise of
$72~\textrm{fT/(cm} \sqrt{\textrm{Hz}})$ for the white noise level
and $880~\textrm{fT/(cm} \sqrt{\textrm{Hz}})$ at 1 Hz in unshielded
environments. We are continuing to investigate the stability over
longer time periods and include the noise investigations in shielded as well as unshielded environment.

\section*{Acknowledgment}
The authors would like to thank L. Redlich (IPHT Jena) for laser
patterning of the grain boundary markers of the samples. This work
is partially supported by EU BIODIAGNOSTICS (017002).

\bibliography{literature}{}

\end{document}